\documentclass[11pt]{article}
\usepackage[dvips]{graphics}
\usepackage{graphicx}
\title{Probability in relativistic Bohmian mechanics of particles and strings}
\author{Hrvoje Nikoli\'c \\
Theoretical Physics Division, Rudjer Bo\v{s}kovi\'{c} Institute, \\
P.O.B. 180, HR-10002 Zagreb, Croatia \\
{\normalsize e-mail: hrvoje@thphys.irb.hr} \\
\makebox[1in]{} \\
}
\date{\today}
\begin{document}
\maketitle
\begin{abstract}
Even though the Bohmian trajectories given by integral curves of the
conserved Klein-Gordon current may involve motions backwards in time,
the natural relativistic probability density of particle positions is well-defined.
The Bohmian theory predicts
subtle deviations from the statistical predictions
of more conventional formulations of quantum theory,
but it seems that no present experiment rules this theory out.
The generalization to the case of many particles or strings 
is straightforward, provided that a preferred 
foliation of spacetime is given. 
\end{abstract}
\vspace*{0.5cm}
\maketitle

\section{Introduction}

The Bohmian interpretation of nonrelativistic quantum mechanics (QM) 
\cite{bohm12,bohmrep1,holbook} is the best
known and most successfull reformulation of QM in terms of hidden variables.
However, the generalization to the relativistic case is still an unsettled issue.
One of the approaches is 
the most direct generalization based on 3 natural steps:
(i) the nonrelativistic Schr\"odinger equation is replaced by
the corresponding relativistic wave equation, (ii) the 
conserved probability current (associated with the Schr\"odinger equation) 
is replaced by the appropriate relativistic conserved current associated with the 
relativistic wave equation, 
and (iii) the relativistic Bohmian particle trajectories are postulated to be
the integral curves of this relativistic conserved current.
However, the simplest generalization consisting of these 3 natural steps
is not without difficulties. For example, in \cite{holbook} it has been considered
problematic because for bosonic particles such Bohmian velocities
may be superluminal. Nevertheless, by using the theory of quantum measurements,
it has been stressed that measured velocities cannot be
superluminal \cite{nikfpl1,nikfpl3}, which avoids possible clashes with observations.
Similarly, such Bohmian particles can move backwards in time leading to multiple
particle positions at a single time, but again, by using the theory of quantum measurements
it has been argued that such multiple positions cannot be observed \cite{nikfpl3}.

Another objection against such a version of the Bohmian interpretation
is based on the fact that the time component of the conserved current
may be negative, which means that it cannot be interpreted as the probability density
of particle positions. In \cite{nikfpl3} it was argued that it does not necessarily make
the Bohmian interpretation inconsistent, because at the fundamental 
level Bohmian mechanics is a fully deterministic theory, so it does not need to have an {\it a priori} defined probability density. Instead, the results in nonrelativistic QM \cite{val}
suggest that a simple relation between wave function and probability density may be 
an emergent phenomenon, not a fundamental one. Nevertheless, the absence 
of an {\it a priori} probability density of particle positions makes the theory less
predictive, so even if it is not a fundamental problem, it makes
the theory less useful in practice.

In this paper we show that the natural conserved probability density {\em can} be introduced.
Essentially, it is the absolute value of the time-component of the conserved current, but
in the single-particle case it is fully relativistic-covariant. In the $n$-particle case
it requires a preferred foliation of spacetime, but the foliation can be specified by a 
unit vector normal to the preferred hypersurfaces which allows to write all the equations 
in a relativistic-covariant form.
It can also be further generalized to the case of strings.
The nontrivial aspect of this probability density stems from motions backwards in time and
multiple particle positions, implying that a hypersurface on which the
total probability is equal to one may not be spacelike everywhere or may be given
by only a part of a spacelike Cauchy hypersurface. Nevertheless, such hypersurfaces
are defined by the congruence of the integral curves of the conserved current,
so in principle all statistical predictions are uniquely defined (up to the choice of preferred foliation). As demonstrated in \cite{nikfpl3}, the measurable predictions
may differ from those of more conventional formulations of quantum theory, 
but we argue that no present experiment rules this Bohmian theory out. 
 
The next section deals with the single-particle case, while Secs.~\ref{SEC3} and
\ref{SEC4} contain the generalizations to the cases of many particles and strings,
respectively. The conclusions are drawn in Sec.~\ref{SEC5}.
We use the units $\hbar=c=1$ and the metric signature $(+,-,-,-)$.

\section{Probability in the single-particle case}
\label{SEC2}

\subsection{General theory}

For the sake of brevity, this subsection is {\em not} intended to be self-contained.
Instead, we extensively use some mathematical results and geometrical
insights explained in more detail in
\cite{nikfpl3} and \cite{nikijmpa}. The main line of reasoning can be followed without
explicit reference to these papers, but for the sake of more complete understanding
we recommend to consult these papers as well.

Let $\hat{\phi}(x)$ be a scalar hermitian field operator satisfying the free Klein-Gordon
equation. If $|0\rangle$ is the vacuum and
$|1\rangle$ is an arbitrary 1-particle state, the corresponding c-number
valued wave function is $\psi(x)=\langle 0|\hat{\phi}(x)|1\rangle$ (see also the Appendix).
Such $\psi(x)$ is a superposition of positive-frequency solutions of the Klein-Gordon equation
\begin{equation}
 (\partial^{\mu}\partial_{\mu}+m^2)\psi(x)=0.
\end{equation}
The naturally associated conserved current is the Klein-Gordon current
\begin{equation}\label{cur}
j_{\mu}=i\psi^*\!\stackrel{\leftrightarrow\;}{\partial_{\mu}}\! \psi .
\end{equation}
It is normalized so that
\begin{equation}\label{globcons}
\int_{\Sigma} dS^{\mu}j_{\mu} =1,
\end{equation}
where $\Sigma$ is an arbitrary 3-dimensional spacelike Cauchy hypersurface,
%\begin{equation}
$
dS^{\mu}=d^3x |g^{(3)}|^{1/2} n^{\mu}
$
%\end{equation}
is the covariant measure of the 3-volume on $\Sigma$,
$n^{\mu}$ is the unit future-oriented vector normal to $\Sigma$, 
and $g^{(3)}$ is the determinant of the induced metric on $\Sigma$.
 The conservation equation
\begin{equation}\label{cons}
\partial_{\mu}j^{\mu}=0 
\end{equation}
provides that (\ref{globcons}) does not depend on $\Sigma$.

In the Bohmian interpretation, the particles have trajectories given by integral curves
of the conserved vector field $j^{\mu}(x)$. These curves can be parametrized
by an auxiliary affine scalar parameter $s$, so that the explicit Bohmian equation of motion
reads
\begin{equation}\label{bohmeqm}
\frac{d x^{\mu}(s)}{ds}=j^{\mu} ,
\end{equation}
which determines the trajectory $x^{\mu}(s)$.

If $j^0(x)$ is non-negative for all $x$, then (\ref{cons}) implies that
$j^0(x)$ is the natural conserved probability density of particle positions at a hypersurface
of constant $x^0$. Such a probability density is compatible
with the Bohmian equation of motion (\ref{bohmeqm}). More generally, if 
\begin{equation}\label{jtil}
\tilde{j} \equiv |g^{(3)}|^{1/2} n^{\mu} j_{\mu} 
\end{equation}
is non-negative for all $x$ for every timelike future-oriented $n^{\mu}$,
then $\tilde{j}(x)$ is the natural conserved probability density of particle positions 
at an arbitrary spacelike hypersurface specified by its unit normal $n^{\mu}(x)$.
The results of \cite{nikijmpa} show that this
relativistic-covariant definition of particle density is
also compatible with (\ref{bohmeqm}). 

The non-trivial issue is to generalize this to the case in which $j^0$, or more 
generally $\tilde{j}$, may be negative at some $x$. Nevertheless, the 
generalization is rather simple; to make the probability density non-negative,
one simply has to take the absolute value of $j^0$ \cite{tumdis} or $\tilde{j}$.
Indeed, the absolute value $|j^0|$ also satisfies a {\em local} conservation
equation of the form 
$\partial_0|j^0|  \pm\partial_{i}j^{i}=0$, where the upper (lower) sign
is valid for $x$ at which $j^0$ is positive (negative).
Thus, in general, the local conserved probability density is simply
\begin{equation}\label{density}
 \tilde{p}(x)=|\tilde{j}(x)| .
\end{equation}

The non-trivial aspect of (\ref{density}) is the correct {\em global} interpretation
of it. In general, from (\ref{globcons}) we see that
\begin{equation}\label{globcons2}
\int_{\Sigma} d^3x |\tilde{p}| =
\int_{\Sigma} dS^{\mu}|j_{\mu}|  \, \geq 1,
\end{equation}
while, according to the standard theory of probability,
the sum of probabilities for all possibilities of particle positions 
that constitute the sample space
should be strictly equal to 1, not $\geq 1$. Nevertheless, the interpretation of this apparent
inconsistency is rather simple: owing to the deterministic motions of particles
described by (\ref{bohmeqm}), not all possibilities of particle positions on $\Sigma$ 
count as different elements of the sample space.
More precisely, if a trajectory crosses $\Sigma$ at a point $x_A$, 
then the same trajectory may cross the same $\Sigma$ at another point $x_B$,
in which case $x_A$ and $x_B$ represent the same element of the sample
space.
Indeed, as discussed in more detail in \cite{nikfpl1,nikfpl3}, and mathematically
more rigorously in \cite{tumdis}, this is a direct
consequence of the fact that $j^{\mu}$ may be spacelike and $j^0$ may be negative
at some regions of spacetime. Therefore, instead of using the timelike
Cauchy hypersurface $\Sigma$ in (\ref{globcons2}), one must use a different
3-dimensional hypersurface $\Sigma'$, chosen such that {\em no trajectory crosses $\Sigma'$ more than ones}. On such a hypersurface one has
\begin{equation}\label{globcons3}
\int_{\Sigma'} d^3x |\tilde{p}| \, \leq 1.
\end{equation}
If $\Sigma'$ is such that (\ref{globcons3}) 
is equal to 1, then we say that $\Sigma'$ is {\em complete}.
The case $<1$ in (\ref{globcons3}) may occur because $\Sigma'$ may be chosen such that
some trajectories never cross $\Sigma'$. Typically, the case $<1$ corresponds to 
the case in which $\Sigma'$ is a subset of the Cauchy hypersurface $\Sigma$ \cite{nikfpl3}.

In practice, it is generally nontrivial to find 
a complete $\Sigma'$. This is because, in principle, one needs to know the whole
congruence of integral curves of the vector field $j^{\mu}(x)$, i.e., all the trajectories in the whole
spacetime. Nevertheless, for given $\psi(x)$ this is well defined in principle
(and straightforward to find numerically).
There are two typical shapes that such complete hypersurfaces $\Sigma'$
may take. First, one may require that $\Sigma'$ should be connected.
In this case some regions of $\Sigma'$ may not be spacelike \cite{nikijmpa}.
Second, one may require that $\Sigma'$ should be spacelike everywhere. 
In this case $\Sigma'$ may not be connected \cite{nikfpl3}.
(Such a disconnected $\Sigma'$ consists of 2 or more mutually disconnected
pieces, each being a connected subset of $\Sigma$.)
Of course, a mixture of these two typical shapes, i.e., a hypersurface
which is neither connected nor spacelike everywhere, is also possible.

It is also of interest to know about the cases in which the trajectories
do not necessarily need to be calculated explicitly. From the results of
\cite{nikfpl3} one can infer the following: If $\Sigma'$ is a connected 
subspace of the spacelike Cauchy hypersurface $\Sigma$ such that
$\tilde{j}$ has the same sign everywhere on $\Sigma'$, then no trajectory
crosses $\Sigma'$ more than ones. Consequently, such 
connected spacelike $\Sigma'$ with constant sign of $\tilde{j}$
can be used in (\ref{globcons3}).

So far we have been tacitly assuming that $\psi$ satisfies the free Klein-Gordon
equation everywhere and that $j^{\mu}(x)$ is a regular vector field everywhere.
However, when the detection process or the initial creation of particles is taken 
into account, this does not longer need to be the case. In particular, 
owing to the motions backwards in time, not all 
trajectories need to cross (not even ones)
every spacelike Cauchy hypersurface $\Sigma$.
For example, it may happen that the particle created at $t_0$ never
reaches the detector starting with operation at $t_1>t_0$. Another possibility
is that some trajectories become completely unphysical, in the 
sense that no initial condition corresponding to a particle existing at $t_0$
is compatible with these trajectories \cite{nikfpl3}. This leads to an interesting
prediction that, instead of multiple particle positions, one will actually 
observe that a particle will never be found at some 
positions at which the wave function does not vanish \cite{nikfpl3}.
(In such cases, the probability density on different regions of spacetime 
is either given by (\ref{density}) or equal to zero, which,
as demonstrated in \cite{nikfpl3}, is also determined by the whole
congruence of integral curves induced by $j^{\mu}(x)$.)

%Nevertheless, given the practical limitations we discussed above, it is not clear
%if such deviations from the predictions of more conventional formulations of quantum
%theory can be observed with current technology. 

\subsection{On measurable consequences}

As we have seen, negative values of $j_0$ are related to motions backwards in time,
which may lead to multiple particle positions at the same time. This, of course,
can be interpreted as particle creation. However, since this occurs
even for free particles, such a prediction of particle creation does not coincide
with predictions on particle creation in more conventional formulations of quantum theory. 
(Note, however, that {\em the} standard prediction of probabilities of 
particle positions does not really exist, because the standard approach
requires a relativistic position operator, which does not exist.
Therefore, the issue of probabilities of relativistic particle positions is an
unsettled issue even within the conventional formulations of quantum theory \cite{nikmyth}.)
Owing to the difference between the predictions of the Bohmian 
and the conventional formulation,
one could jump to the conclusion that this makes such a Bohmian interpretation
untenable. A more optimistic view is that this difference 
could be used to test the Bohmian formulation
experimentally. However, in this subsection we argue that it is actually rather
difficult to see the differences in practice and that probably no currently existing
experiment can be used to rule out such a version of the Bohmian interpretation.

First, most experiments on relativistic particles are based on scattering experiments.
Such experiments are better viewed as measurements of particle momenta
(rather than positions), in which case the predictions of the Bohmian interpretation
coincide with those of the standard interpretation \cite{nikfpl1}.

Next, experiments that measure quantum probabilities of particle positions (e.g. measurements
of interference patterns) do exist, but they are usually based on
stationary wave functions, namely functions of the form
\begin{equation}
 \psi(x)=\frac{e^{-i\omega t}}{\sqrt{2\omega}} \varphi({\bf x}) .
\end{equation}
For such wave functions
\begin{equation}\label{11}
 j_0({\bf x},t)=|\varphi({\bf x})|^2 ,
\end{equation}
so the probability density of particle positions coincides with that of 
a more conventional formulation. 
More generally, whenever $j_0$ is non-negative everywhere
one may view $j_0$ as a conventional probability density, because
in conventional views $j_0$ can be identified with charge density,
so for a neutral particle it is reasonable to expect that it coincides with 
the probability density.

Thus, to obtain a prediction of the Bohmian interpretation that differs from conventional ones,
we must deal with a case in which $j_0$ may be negative.
The necessary (though not sufficient) condition is that the state should be a superposition
of two or more different frequencies. Therefore, let us study the case of two
different equally probable frequencies
\begin{equation}
 |1\rangle = \frac{|k_1\rangle +|k_2\rangle}{\sqrt{2}} , 
\end{equation}
where $|k_1\rangle$ and $|k_2\rangle$ are the 4-momentum eigenstates
with 4-momenta $k_1$ and $k_2$, respectively. The corresponding wave
function (normalized in a finite 3-volume $V$) is
\begin{equation}\label{13}
 \psi(x)=\frac{1}{\sqrt{2}} \left[ \frac{e^{-ik_1\cdot x}}{\sqrt{V2\omega_1}}
+ \frac{e^{-ik_2\cdot x}}{\sqrt{V2\omega_2}} \right] ,
\end{equation}
where $\omega_{1,2}=\sqrt{{\bf k}_{1,2}^2+m^2}$. 
Thus (\ref{cur}) gives
 \begin{equation}\label{jconc}
  j^{\mu}(x)=\frac{1}{2V} \left[ \frac{k_1^{\mu}}{\omega_1} +
\frac{k_2^{\mu}}{\omega_2} + 
\frac{k_1^{\mu}+k_2^{\mu}}{\sqrt{\omega_1\omega_2}} \, {\rm cos}[(k_1-k_2)\cdot x]
\right] .
 \end{equation}
We know that the non-relativistic limit leads to non-negative $j^0$, so the most
interesting case is expected to be the ultrarelativistic limit $m\rightarrow 0$.
Therefore, we study the case $m=0$. For simplicity, we study the
case of $1+1$ dimensional motion, i.e., we assume that $k_{1,2}^{\mu}$
is nonvanishing only for $\mu=0,1$. Thus, the momenta $k_1$ and $k_2$
are either collinear (the space components of momenta have the same directions)
or anti-collinear  (the space components of momenta have the opposite directions).

First consider the case of collinear momenta. Thus we take $k_1^1=\omega_1$,
$k_2^1=\omega_2$ (the upper label means $\mu=1$), so
the non-vanishing components of
(\ref{jconc}) are
 \begin{equation}\label{jconc0}
 j^0=\frac{1}{V} \left[ 1+ 
\frac{\omega_1+\omega_2}{2\sqrt{\omega_1\omega_2}} \, {\rm cos}
[(\omega_1-\omega_2)(t-x^1)]
\right] ,
 \end{equation}
\begin{equation}\label{jconc1}
 j^1=\frac{1}{V} \left[ 1+ 
\frac{\omega_1+\omega_2}{2\sqrt{\omega_1\omega_2}} \, {\rm cos}
[(\omega_1-\omega_2)(t-x^1)]
\right] .
 \end{equation}
We see that (\ref{jconc0}) is negative for some $x$, provided that $\omega_1\neq\omega_2$. 
Nevertheless, we see that $j^1= j^0$, which means
that the trajectories satisfy $dx^1/dx^0=j^1/j^0=1$, i.e.
all trajectories have a constant velocity equal to the velocity of light.
In other words, even though $j^0$ is negative for some $x$, there are no motions
backwards in time and thus there are no measurable deviations from more conventional 
formulations of quantum theory.

Now consider the case of anti-collinear momenta. Thus we take $k_1^1=\omega_1$,
$k_2^1=-\omega_2$, which leads to
\begin{equation}\label{jconc0a}
 j^0=\frac{1}{V} \left[ 1+ 
\frac{1+\eta}{2\sqrt{\eta}} \, {\rm cos}[(1-\eta)\omega_1t-(1+\eta)\omega_1x^1] 
\right] ,
\end{equation}
\begin{equation}\label{jconc1a}
 j^1=\frac{1}{V} \left[ 
\frac{1-\eta}{2\sqrt{\eta}} \, {\rm cos}[(1-\eta)\omega_1t-(1+\eta)\omega_1x^1] 
\right] ,
 \end{equation}
where $\eta\equiv \omega_2/\omega_1$. Now $j^0$ is negative for some $x$
and $j^0\neq j^1$, so motions backwards in time are possible. (We have confirmed that
by numerically finding the trajectories determined by (\ref{jconc0a})-(\ref{jconc1a}).)
However, in practice, it seems to be very difficult to measure particle positions 
for such a state. Namely, this state corresponds to
a superposition of two coherent beams moving in the opposite directions, so they cannot both 
hit the detection screen from the same side. Thus, owing to the
anti-collinear momenta, the effects of interference
cannot be seen on the screen.

To overcome this problem, one could work with beams that are neither collinear
nor anti-collinear. For example, one could do a variant of the two-slit experiment
in which slit 1 transmits a wave with frequency $\omega_1$,
while slit 2 transmits a wave with frequency $\omega_2$.
So let us generalize the analysis above to incorporate such possibilities as well.
In a conventional approach, one deals with a wave function of the form
(see the Appendix)
\begin{equation}\label{wf1}
 \varphi({\bf x},t)=\frac{1}{\sqrt{2}} [ e^{-i\omega_1t} \varphi_1({\bf x})
+ e^{-i\omega_2t} \varphi_2({\bf x}) ] ,
\end{equation}
where $\varphi_{1,2}({\bf x})$ are normalized such that
$\int d^3x\, |\varphi_{1,2}({\bf x})|^2 =1$. The 
associated conventional probability density (see the Appendix)
is $\rho({\bf x},t)=|\varphi({\bf x},t)|^2$. This gives 
\begin{equation}\label{int1}
 \rho({\bf x},t)=C({\bf x})+I({\bf x},t) ,
\end{equation}
where 
\begin{equation}
 C({\bf x})=\frac{1}{2} [|\varphi_1({\bf x})|^2 + |\varphi_2({\bf x})|^2 ] 
\end{equation}
is the ``classical'' probability density and
\begin{equation}
 I({\bf x},t)=\frac{1}{2} [ e^{-i(\omega_1-\omega_2)t}
\varphi_1({\bf x})\varphi_2^*({\bf x}) + e^{i(\omega_1-\omega_2)t}
\varphi_1^*({\bf x})\varphi_2({\bf x}) ]
\end{equation}
is the interference term.

In the approach based on the Klein-Gordon 
scalar product (see the Appendix), instead of (\ref{wf1}) 
one deals with a differently normalized wave function 
\begin{equation}\label{wf2}
 \psi({\bf x},t)=\frac{1}{\sqrt{2}} \left[ \frac{e^{-i\omega_1t}}{\sqrt{2\omega_1}}
\varphi_1({\bf x})
+ \frac{e^{-i\omega_2t}}{\sqrt{2\omega_2}}
\varphi_2({\bf x}) \right] ,
\end{equation}
which generalizes (\ref{13}).
This leads to
\begin{equation}\label{int2}
 j_0({\bf x},t)=C({\bf x})+\alpha I({\bf x},t) ,
\end{equation}
where
\begin{equation}
 \alpha\equiv \frac{\omega_1+\omega_2}{2\sqrt{\omega_1\omega_2}} .
\end{equation}

In the limit $\omega_1=\omega_2$ we have $\alpha=1$, so
from (\ref{int1}) and (\ref{int2}) one recovers (\ref{11})
\begin{equation}
 \rho=j_0=C({\bf x})+I({\bf x}) ,
\end{equation}
where $ I({\bf x})=[ \varphi_1({\bf x})\varphi_2^*({\bf x}) + 
\varphi_1^*({\bf x})\varphi_2({\bf x}) ]/2$. This describes a usual
stationary interference pattern. When $\omega_1\neq\omega_2$,
then both (\ref{int1}) and (\ref{int2}) describe a nonstationary
interference pattern that oscillates with the frequency $\omega_1-\omega_2$.
Since $\alpha\neq 1$, these two patterns are different, which, in principle, 
could be distinguished experimentally. However, if $|\omega_1-\omega_2|$
is large so that the oscillation is too fast to see it experimentally, then everything
that can be seen is the time-averaged distribution
\begin{equation}
 \langle\rho\rangle=\langle j_0\rangle=C({\bf x}) ,
\end{equation}
that washes out all effects of interference and all differences
between the two approaches. 
(More precisely, the time average $\langle j_0\rangle=C({\bf x})$ is 
observable if $j_0$ is non-negative. If it is negative at some regions 
then one should actually calculate $\langle |j_0|\rangle$ which may differ
from $C({\bf x})$. However, the typical distances at which the deviations
from $C({\bf x})$ occur are of the order $|\omega_1-\omega_2|^{-1}$,
which are small when $|\omega_1-\omega_2|$ is large.)
To see the oscillations one must have 
small $|\omega_1-\omega_2|$ (i.e. $\omega_1\simeq\omega_2$), 
but then $\alpha\simeq 1$
so the differences between (\ref{int1}) and (\ref{int2}) are 
difficult to see again. 

\section{Generalization to the many-particle case}
\label{SEC3}

Now we generalize
the single-particle wave function $\psi(x)$ to the 
$n$-particle wave function $\psi(x_1,\ldots,x_n)$ \cite{nikfpl3}.
It satisfies $n$ Klein-Gordon equations, one for each $x_a$,
$a=1,\ldots,n$. 
Similarly, there are $n$ conserved Klein-Gordon currents
\begin{equation}\label{curn}
j_{a\mu}=i\psi^*\!\stackrel{\leftrightarrow\;\;\;}{\partial_{a\mu}}\! \psi ,
\end{equation}
\begin{equation}\label{consn}
\partial_{a\mu}j_a^{\mu}=0 .
\end{equation}
Eq.~(\ref{consn}) is valid for each $a$, but these equations can also be summed to give
\begin{equation}\label{consn2}
\sum_a \partial_{a\mu}j_a^{\mu}=0 . 
\end{equation}
In the Bohmian interpretation one postulates \cite{durr96,nikfpl3,niktalk}
\begin{equation}\label{bohmeqmn}
\frac{d x_a^{\mu}(s)}{ds}=j_a^{\mu} ,
\end{equation}
which determines $n$ trajectories $x_a^{\mu}(s)$.
These $n$ trajectories in the 4-dimensional spacetime can also be viewed as one trajectory 
in the $4n$-dimensional configuration spacetime. 
%Instead of the usual Lorentz symmetry
%SO(1,3), the configuration spacetime is characterized by the SO($n$,$3n$) symmetry.
%(For a discussion of the configuration spacetime see also \cite{pavs}.)

Even though the Bohmian equation of motion (\ref{bohmeqmn}) for $n$ particles
is nonlocal, it is completely relativistic covariant \cite{niktalk}. 
No {\it a priori} preferred foliation of spacetime is required. 
The functions $x_a^{\mu}(s)$ can be determined by a specification of $4n$ ``initial'' 
conditions $x_a^{\mu}(0)$. 
However, the price payed for this large symmetry is a smaller predictive power. 
Various choices of these ``initial'' conditions correspond to 
various choices of synchronization among the $n$ particles \cite{niktalk}.  

To increase the predictive power of the theory, in the following we consider
a different version of the theory, a version with a smaller symmetry. 
The Lorentz symmetry brakes by introducing a preferred foliation of spacetime
specified by the timelike future-oriented unit normal vector $N^{\mu}(x)$.
It satisfies $\nabla_{\mu}N^{\mu}=0$, where $\nabla_{\mu}$ is the covariant
derivative generalizing the ordinary derivative $\partial_{\mu}$ to curved coordinates.
Following \cite{nikijmpa}, we introduce the $n$-vector
\begin{equation}\label{curnold}
j_{\mu_1\ldots\mu_n}(x_1,\ldots,x_n)=i^n \psi^*
\!\stackrel{\leftrightarrow}{\partial}_{\mu_1}\! \cdots
\!\stackrel{\leftrightarrow}{\partial}_{\mu_n}\! \psi ,
\end{equation}
where $\partial_{\mu_a}\equiv \partial/\partial x^{\mu_a}_a$.
Now, analogously to the fermionic case studied in \cite{durr99},
we introduce $n$ currents $j_{\mu_a}(x_1,\ldots,x_n)$,
$a=1,\ldots,n$, by contracting
(\ref{curnold}) $(n-1)$ times with the vector $N^{\mu}$. For example,
for $a=1$,
\begin{equation}\label{curn1}
 j_{\mu_1}(x_1,\ldots,x_n)=j_{\mu_1\ldots\mu_n}(x_1,\ldots,x_n)
N^{\mu_2}(x_2)\cdots N^{\mu_n}(x_n) ,
\end{equation}
which satisfies $\nabla_{\mu_1} j^{\mu_1}=0$. 
In (\ref{curn1}) it is understood that all points lie at the same hypersurface 
of the preferred foliation.
Thus, instead of (\ref{bohmeqmn}),
now the Bohmian particle trajectories are postulated to be
\begin{equation}\label{bohmeqmn2}
\frac{d x_a^{\mu}(s)}{ds}=j_{\mu_a} .
\end{equation} 
From the results of \cite{durr99} and \cite{nikijmpa} one finds that the
probability density of particle positions on preferred hypersurfaces is
\begin{equation}\label{densityn}
\tilde{p}(x_1,\ldots,x_n) = |\tilde{N}^{\mu_1}(x_1) \cdots
\tilde{N}^{\mu_n}(x_n) j_{\mu_1\ldots\mu_n}(x_1,\ldots,x_n)| ,
\end{equation}
where
\begin{equation}
\tilde{N}^{\mu_a}(x_a)=|g^{(3)}_a(x_a)|^{1/2} N^{\mu_a}(x_a).
\end{equation} 

\section{Generalization to strings}
\label{SEC4}

The Bohmian interpretation of string theory has been studied in 
\cite{nikstr1,nikstr2,nikstr3,nikstr4}.
In the Bohmian context, strings have several advantages over particles
or fields.
First, bosons and fermions are treated symmetrically \cite{nikstr3}.
Second, the symmetry between bosons and fermions provides new 
insights on the origin of preferred foliation of spacetime at the level
of effective field theory \cite{nikstr3}. 
Third, the Bohmian equation of motion for strings automatically includes
a continuous description of particle creation and destruction \cite{nikstr3,nikstr4}.
(By contrast, to make Bohmian mechanics of particles compatible
with particle creation and destruction, one is forced to make 
some artificial modifications of the theory \cite{durr1,durr2}, \cite{nikfpl1,nikfpl2}.)

Essentially, strings are obtained from many-particle systems through a replacement
of the discrete label $a$ by a continuous variable $\sigma$. 
Thus, instead of $n$ coordinates $x^{\mu}_a$ we deal with a continuous set 
of string coordinates $x^{\mu}(\sigma)$.
In 
\cite{nikstr1,nikstr2,nikstr3,nikstr4} we have studied the spacetime covariant
version, specified by the wave functional $\psi[x(\sigma)]\equiv\psi[x]$. 
Here, by analogy with 
Sec.~\ref{SEC3}, we introduce a preferred foliation of spacetime 
specified by $N^{\mu}(x)$.
(Of course, in string theory the number of space dimensions is not 3, but
25 in bosonic string theory and 9 in superstring theory \cite{GSW,polc}.)
Next, we introduce the local symmetric hermitian functional-derivative operator
\begin{equation}
 \hat{P}(\sigma)=i\left[ 
N^{\mu}(x(\sigma)) \frac{ \stackrel{\rightarrow}\delta } {\delta x^{\mu}(\sigma)} -
\frac{ \stackrel{\leftarrow}{\delta} } {\delta x^{\mu}(\sigma)} N^{\mu}(x(\sigma)) \right] .
\end{equation}
Now, for bosonic strings, the string current is given by a generalization of (\ref{curn1})  
\begin{equation}\label{strc}
j_{\mu}[x;\sigma)=i\psi^*[x]
\frac{
\!\stackrel{\leftrightarrow}{\delta}\! }
{\delta x^{\mu}(\sigma)}
\left\{ \prod^{(\sigma)}_{\sigma'}\hat{P}(\sigma') \right\}
\psi[x] .
\end{equation}
Here the notation $[x;\sigma)$ denotes a functional with respect to $x$ and a function
with respect to $\sigma$, while the product $\prod^{(\sigma)}_{\sigma'}$ denotes the
product over all values of $\sigma'$ except $\sigma'=\sigma$.
In the superstring case it generalizes to 
\begin{equation}\label{strcs}
j_{\mu}[x;\sigma)=i  \int [dM] \, \psi^*[x,M]
\frac{
\!\stackrel{\leftrightarrow}{\delta}\! }
{\delta x^{\mu}(\sigma)}
\left\{ \prod^{(\sigma)}_{\sigma'}\hat{P}(\sigma') \right\}
\psi[x,M] .
\end{equation}
where $M(\sigma)$ is an additional variable generalizing the spinor indices 
of particle wave functions \cite{nikstr3}. Note that bosonic and fermionic string states
are described by a single universal current (\ref{strcs}), which is a 
generalization of the Klein-Gordon (not the Dirac) current. Consequently, 
if superstring theory is correct, then, in the particle limit, the Bohmian particle trajectories
of fermions are also described by a version of the Klein-Gordon current \cite{nikstr3,nikstr4}.
The string current is conserved
\begin{equation}\label{conss}
\int d\sigma \,  \frac{ \delta j^{\mu}[x;\sigma)}{\delta  x^{\mu}(\sigma)} =0 ,
\end{equation}
which implies that the local probability density that the string has the shape 
$x(\sigma)$ is given by a generalization of (\ref{densityn})
\begin{eqnarray}
 \tilde{p}[x] & = & \left| \left\{ \prod_{\sigma'}|g^{(3)}(x(\sigma'))|^{1/2} \right\} 
N^{\mu}(x(\sigma)) \,  j_{\mu}[x;\sigma) \right| 
\nonumber \\
& = & \left| \int [dM] \, \psi^*[x,M] 
\left\{ \prod_{\sigma'}\hat{\tilde{P}}(\sigma') \right\}
\psi[x,M] \right| ,
\end{eqnarray}
where $\prod_{\sigma'}$ denotes the product over all values of $\sigma'$
and $\hat{\tilde{P}}$ is obtained from $\hat{P}$ by a replacement
$N^{\mu}\rightarrow \tilde N^{\mu}$.
This probability density is consistent with the Bohmian trajectories described by the functions
$x^{\mu}(\sigma,s)$, satisfying the Bohmian equation of motion
\begin{equation}\label{bohmeqmn4}
\frac{d x^{\mu}(\sigma,s)}{ds}=j^{\mu}[x;\sigma) ,
\end{equation}
which generalizes (\ref{bohmeqmn2}).

\section{Conclusions}
\label{SEC5}

Even though the time component $j_0({\bf x},t)$ of the conserved Klein-Gordon current 
is not positive definite, the absolute value $|j_0({\bf x},t)|$ is. Therefore, 
as we have shown, this absolute value
is the natural probability density of particle positions at time $t$. 
Further, we have shown that the fully relativistic
covariant generalization of this is given by the probability density (\ref{density}). 
Indeed, such probability density is locally conserved. The issue of global probability
conservation is less trivial, but we have seen that the knowledge of the whole congruence 
of all Bohmian trajectories settles this issue as well.
These results show that the Bohmian interpretation based on the Klein-Gordon current
is fully relativistic covariant and fully predictive.

Further, although in some cases the predictions of this version of Bohmian mechanics may
differ from the predictions of more conventional approaches to quantum theory,
we have demonstrated that in practice such differences are difficult to observe.
It seems that no already done experiment can be used to rule out this version
of the Bohmian interpretation.
Nevertheless, our results on practical measurability are not yet conclusive, so we 
challenge the readers to find an achievable experimental test that could
confirm or falsify the predictions of this theory.

Finally, we have generalized our results to many-particle systems and strings.
In agreement with \cite{durr96,durr99}, we have found that a fully predictive theory
with well-defined probabilities of particle positions cannot be constructed
in a fully relativistic manner. Nevertheless, by introducing a preferred foliation of spacetime
specified by the vector field of unit normals to hypersurfaces of the foliation, 
all equations can be written in a relativistic-covariant form.
%Nevertheless, our approach is still relativistic
%covariant with respect to a configuration spacetime. The physical
%relevance of this configuration relativistic covariance is still obscure, 
%but we hope that this could motivate further research (see also \cite{tum,pavs}).

\section*{Acknowledgements}

The author is grateful to R.~Tumulka for valuable discussions and suggestions.
This work was supported by the Ministry of Science of the
Republic of Croatia under Contract No.~098-0982930-2864.

\appendix
\section{Relativistic wave functions and their normalizations}

A free scalar hermitian field operator can be expanded as \cite{ryder}
\begin{equation}\label{a1}
 \hat{\phi}(x)=\int \frac{d^3k}{(2\pi)^3 2\omega_{\bf k}} \,
[\hat{a}({\bf k})e^{-ik\cdot x} + \hat{a}^{\dagger}({\bf k})e^{ik\cdot x}] ,
\end{equation}
where $k^{\mu}=(\omega_{\bf k},{\bf k})$ and
$\omega_{\bf k}=\sqrt{{\bf k}^2+m^2}$. 
The destruction and creation operators satisfy
\begin{equation}\label{a2}
 [\hat{a}({\bf k}),\hat{a}^{\dagger}({\bf k}')] =
(2\pi)^3 2\omega_{\bf k} \delta^3({\bf k}-{\bf k}').
\end{equation}
The quantities $d^3k/2\omega_{\bf k}$ and 
$2\omega_{\bf k} \delta^3({\bf k}-{\bf k}')$ are Lorentz invariant \cite{ryder}, 
which shows that the normalizations in (\ref{a1}) and (\ref{a2})
are manifestly Lorentz invariant.

Let $c({\bf q})$ be an arbitrary c-number valued function normalized so that
\begin{equation}\label{a3}
\int \frac{d^3q}{(2\pi)^3} \, |c({\bf q})|^2=1 .
\end{equation}
Such a function can be used to define the most general 1-particle state
\begin{equation}\label{a4}
 |1\rangle = \int \frac{d^3q}{(2\pi)^3} \, c({\bf q}) 
\frac{ \hat{a}^{\dagger}({\bf q}) }{\sqrt{2\omega_{\bf q}}} |0\rangle ,
\end{equation}
where $|0\rangle$ is the vacuum, $\hat{a}({\bf q})|0\rangle =0$.
Using (\ref{a2}) and (\ref{a3}), one finds that the normalization in 
(\ref{a4}) provides that $\langle 1|1\rangle = 1$.
Following \cite{ryder}, the wave function associated with (\ref{a4}) can be defined as
$\psi(x)= \langle 0|\hat{\phi}(x)|1\rangle$. Using (\ref{a1}),  (\ref{a2}) and (\ref{a4}),
this gives
\begin{equation}\label{a5}
 \psi(x)= \int \frac{d^3q}{(2\pi)^3} \, c({\bf q}) 
\frac{ e^{-iq\cdot x}}{\sqrt{2\omega_{\bf q}}} .
\end{equation}
The norm of this wave function is defined through the Klein-Gordon scalar product
\begin{equation}\label{a6}
(\psi,\psi')=i\int_{\Sigma} 
dS^{\mu}\, \psi^*\!\stackrel{\leftrightarrow\;}{\partial_{\mu}}\! \psi' . 
\end{equation}
When $\psi(x)$ and $\psi'(x)$ satisfy the Klein-Gordon equation, then 
(\ref{a6}) does not depend on the choice of the spacelike Cauchy hypersurface
$\Sigma$. Therefore we choose $\Sigma$ to be a hypersurface of constant
Lorentz time-coordinate $x^0$. This implies that the norm of
(\ref{a5}) is
\begin{equation}\label{a7}
 (\psi,\psi)=i\int d^3x\, \psi^*\!\stackrel{\leftrightarrow\;}{\partial_0}\! \psi =1 ,
\end{equation}
where the identity
\begin{equation}\label{a8}
 \int \frac{d^3x}{(2\pi)^3} \, e^{-i({\bf k}-{\bf k}') {\bf x}} = \delta^3({\bf k}-{\bf k}')
\end{equation}
and normalization (\ref{a3}) have been used.

The wave function (\ref{a5}) is not the only meaningful wave function that can
be associated with the state (\ref{a4}). Another possibility is to introduce the wave function
\begin{equation}\label{a9}
 \varphi(x)= \int \frac{d^3q}{(2\pi)^3} \, c({\bf q}) e^{-iq\cdot x} .
\end{equation} 
Using (\ref{a8}) one finds that (\ref{a9}) has the property
\begin{equation}\label{a10}
 \int d^3x \, \varphi^*(x)\varphi(x)=
\int \frac{d^3k}{(2\pi)^3} \int d^3k' \, c^*({\bf k}) c({\bf k}')
e^{i(\omega_{\bf k}-\omega_{{\bf k}'})t} \delta^3({\bf k}-{\bf k}') . 
\end{equation}
At first sight, this quantity seems time-dependent. However,
owing to the $\delta$-function we have ${\bf k}={\bf k}'$. This implies
that $e^{i(\omega_{\bf k}-\omega_{{\bf k}'})t}=1$, which removes
the time-dependence. Consequently, (\ref{a10}) reduces to
\begin{equation}\label{a11}
 \int d^3x \, \varphi^*(x)\varphi(x)=1 ,
\end{equation}
where (\ref{a3}) also has been used. This shows that the integral
in (\ref{a11}) does {\em not} depend on time, i.e. that the quantity
 \begin{equation}\label{a12}
  \rho({\bf x},t)=\varphi^*({\bf x},t)\varphi({\bf x},t) 
 \end{equation}
{\em can} be interpreted as the probability density of particle positions
at time $t$ \cite{jostpuur}. On the other hand, the wave function $\varphi({\bf x},t)$ 
satisfies the Klein-Gordon equation and it is well-known that the
integral on the left-hand side of (\ref{a11}) may depend on time 
when $\varphi$ satisfies the Klein-Gordon equation (which is why
a more complicated scalar product (\ref{a6}) has been introduced in the first place). 
So how is that possible that (\ref{a11}) does not depend on time?
Where is the catch? 
The catch is \cite{jostpuur}
%\footnote{I am grateful to Jouni Puuronen for pointing this out to me.}
that (\ref{a9}) is {\em not} the most general solution of the Klein-Gordon equation.
The most general solution involves a superposition of plane waves with both 
positive and negative frequencies, while (\ref{a9}) contains {\em only positive
frequencies}. Indeed, if both positive and negative frequencies were involved, 
then (\ref{a10}) would also involve factors of the form
$e^{i(\omega_{\bf k}+\omega_{{\bf k}'})t}$ and 
$e^{-i(\omega_{\bf k}+\omega_{{\bf k}'})t}$, which would {\em not} become
time-independent when ${\bf k}=\pm{\bf k}'$.
Thus, the restriction to the space of positive-frequency solutions 
allows us to use the conventional norm (\ref{a11}) and to
interpret (\ref{a12}) as the conserved probability density. 

Nevertheless, the Klein-Gordon norm  (\ref{a7}) still has an advantage
over the conventional norm (\ref{a11}). While (\ref{a7}) is Lorentz
invariant, (\ref{a11}) is {\em not} Lorentz invariant.
This is the main disadvantage of the probability density (\ref{a12}).
Still, in a conventional operational interpretation of QM without hidden variables,
this is not necessarily a problem if one simply postulates that the
``preferred'' Lorentz frame is the frame in which the observer is at rest.
On the other hand, in hidden-variable interpretations in which physical
quantities are assumed to make sense even without observers,
such a subjective identification of the ``preferred'' Lorentz frame
is unacceptable.

\end{document}